\documentclass[twocolumn,amsmath,amssymb,showpacs,superscriptaddress,pra]{revtex4}

\usepackage{graphicx}
\usepackage{dcolumn}
\usepackage{bm}
\usepackage{graphicx,color}

\newcommand{\mb}{\mathbf}

\begin{document}

\title{Hidden momentum in a hydrogen atom and the Lorentz force law}

\author{J. S. Oliveira Filho}\email{juvenil@fisica.ufmg.br}
\affiliation{Departamento de F\'isica, Universidade Federal de Minas Gerais, 30161-970 Belo Horizonte-MG, Brazil}
\affiliation{Universidade Federal do Rec\^oncavo da Bahia, 45300-000 Amargosa-BA, Brazil}
\author{Pablo L. Saldanha}\email{saldanha@fisica.ufmg.br}
\affiliation{Departamento de F\'isica, Universidade Federal de Minas Gerais, 30161-970 Belo Horizonte-MG, Brazil}

\date{\today}

\begin{abstract}
By using perturbation theory, we show that a hydrogen atom with magnetic moment due to the orbital angular momentum of the electron has ``hidden momentum'' in the presence of an external electric field. This means that the atomic electronic cloud has a nonzero linear momentum in its center-of-mass rest frame due to a relativistic effect. This is completely analogous to the hidden momentum that a classical current loop has in the presence of an external electric field. We discuss that this effect is essential for the validity of the Lorentz force law in quantum systems. We also connect our results to the long-standing Abraham-Minkowski debate about the momentum of light in material media.
\end{abstract}


\pacs{03.65.Ta, 03.30.+p, 41.20.-q, 03.65.Ge}


\maketitle

\section{Introduction}

The Lorentz force is one of the fundamental laws of Nature, describing how electromagnetic fields exert force in electric charges. In a recent work, Mansuripur has presented a paradox that seemed to imply that the Lorentz force is incompatible with special relativity and momentum conservation \cite{mansuripur1}. He showed that the torque caused by the Lorentz force on a magnetic dipole in the presence of an electric field was zero in the rest frame of the dipole, but different from zero in other frames.  Such an attack to the foundations of the electromagnetic theory obviously called the attention of the scientific community \cite{cho1}. However, the paradox is solved when the ``hidden momentum'' of the magnetic dipole is taken into account \cite{saldanha1,vanzella1,khorrami1,barnett1,griffiths2}. Hidden momentum is a relativistic effect that may lead a magnetic dipole to carry linear momentum in the presence of an electric field even if it is not moving. In the situation described in Mansuripur paradox, there is a ``hidden angular momentum'' whose time derivative is equal to the torque in each reference frame, so there is no angular acceleration of the dipole and the physical description of the system in the different frames are equivalent \cite{saldanha1}. One important lesson derived from Mansuripur paradox is that the validity of the Lorentz force law is thus conditioned to the existence of hidden momentum.  

Hidden momentum was discovered almost 50 years ago \cite{shockley1,penfieldbook}, but so far there were only classical models for it \cite{shockley1,penfieldbook,coleman1,furry1,vaidman1,hnizdo1,babson1,bjorkquist1,brevik1,griffithsbook,zangwillbook}. In all these models the dipole magnetic moment is the result of classical electric currents. Since the validity of the Lorentz force law is conditioned to the existence of hidden momentum, it is thus legitimate to question if the Lorentz force law is valid in quantum systems such as atoms, since the atomic magnetic moments are not the result of classical currents. In order to solve this question, we use perturbation theory to investigate if a hydrogen atom with magnetic moment due to the orbital angular momentum of the electron in the presence of an external electric field has hidden momentum. We compute the system hidden momentum using two different methods, attesting its existence and the consequent validity of the Lorentz force law in quantum systems. We also discuss how this result influences the long-standing Abraham-Minkowski debate about the momentum of light in material media \cite{pfeifer1}.

\section{Classical hidden momentum}

\begin{figure}
\centering
\includegraphics[height=6cm]{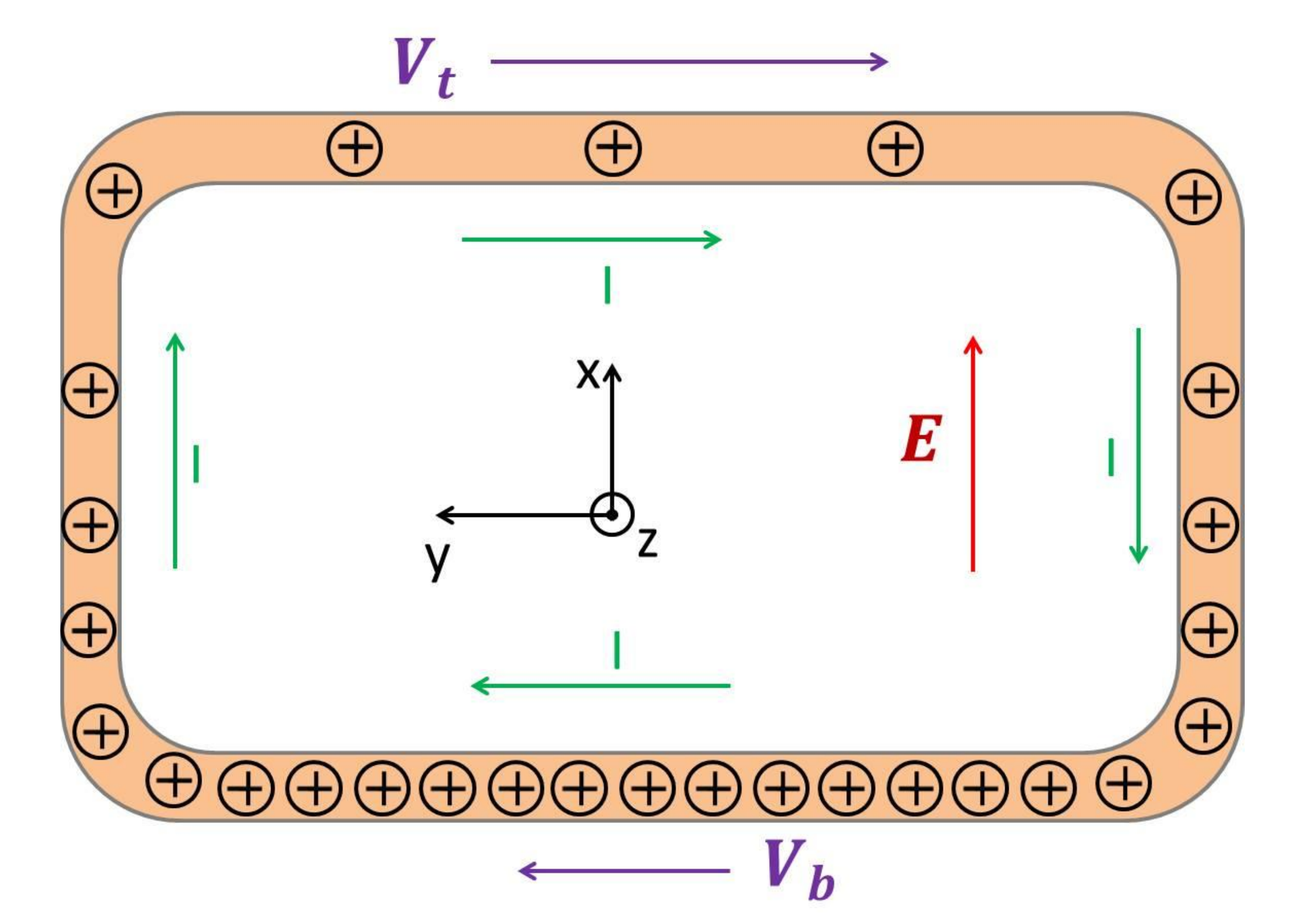}
\caption{A rectangular closed circuit carrying a stationary electric current $I$ in the presence of a constant and uniform external electric field $\mb{E}=E \mb{\hat{x}}$ has hidden momentum. $\mb{V_{t}}$ ($\mb{V_{b}}$) is the velocity of the particles in the top (bottom) segment.}
\label{fig1}
\end{figure}

Before doing our quantum calculations, let us present a brief description of the hidden momentum concept based on a simple classical example \cite{penfieldbook,babson1,  bjorkquist1}. Consider a rectangular closed circuit carrying a stationary electric current generated by a bundle of non-interacting positive charges as depicted in the Fig. \ref{fig1}. The circuit dipole magnetic moment is $\boldsymbol{\mu}=- \mu \mb{\hat{z}}$. When the circuit is in a region with a uniform electric field $\mb{E} = E \mb{\hat{x}}$, the particles on the left (right) segment of the circuit will be accelerated (decelerated) by this field, as shown in the figure. There are more charges moving to the left, but they have a smaller velocity in relation to the particles that move to the right, so if we consider that the momentum of each particle is $M\mb{v}$, where $M$ is the particle mass and $\mb{v}$ its velocity, the total momentum of the particles is zero. But if we consider that the momentum of each particle has the relativistic value $\gamma M\mb{v}$, with $\gamma=1/\sqrt{1-v^2/c^2}$ and $c$ being the speed of light in vacuum, this results in a total relativistic momentum in the $-\mb{\hat{y}}$ direction for the particles given by $\boldsymbol{\mu}\times\mb{E}/c^2$ \cite{penfieldbook,babson1,bjorkquist1}. Since the dipole is not moving, this momentum received the name ``hidden momentum'' \cite{shockley1}. This hidden momentum is counterbalanced by the electromagnetic momentum that results from the integral of the electromagnetic momentum density $\varepsilon_0\mb{E}\times\mb{B}$ in the whole space, where $\mb{B}$ is the magnetic field generated by the dipole and $\varepsilon_0$ is the vacuum permittivity, so the system total momentum is zero \cite{vaidman1,babson1,bjorkquist1}. Several models for magnetic dipoles resulting from classical electric currents predict the existence of hidden momentum \cite{shockley1,penfieldbook,coleman1,furry1,vaidman1,hnizdo1,babson1,bjorkquist1,brevik1,griffithsbook,zangwillbook}. For example, one might prefer to treat the electric current not as a gas of charged particles, but rather as a charged incompressible fluid. In this case the hidden momentum results from a relativistic effect that comes from the difference in pressure between the top and the bottom segments of the circuit \cite{vaidman1,bjorkquist1}. 

There may also be hidden momentum in objects with nonzero electric and magnetic dipole moments even without an external field \cite{babson1,bjorkquist1}. The general expression for hidden momentum in classical systems with static fields can be written as  
\begin{equation}
\mb{P}_{hid} = -\frac{1}{c^{2}} \int{\Phi \mb{J} d^3r},
\label{hidden momentum babson} 
\end{equation}
where $\Phi$ represents the electric potential ($\Phi=-Ex$ in the example of Fig. 1), $\mb{J}$ is the system current density and the volume integral is performed in the whole space \cite{babson1,bjorkquist1}.
 
\section{Hidden momentum in a hydrogen atom}

Now let us proceed to the quantum calculations. In our model for the hydrogen atom, the proton has an infinite mass and is fixed at the origin. The total Hamiltonian of the system formed by the hydrogen atom in the presence of an uniform electric field $\mb{E}=E \mb{\hat{x}}$ can be written as
\begin{equation}
\label{total hamiltonian}
H = H_0+H',\;\mathrm{with}\;H_0=\frac{{p}^{2}}{2 m_{e}}-e V(r) \;\mathrm{and}\; H'= eE x,
\end{equation}
where $\mb{p}$ is the non-relativistic linear momentum operator for the electron, $m_{e}$ its mass, $-e$ its charge, and $V(r)$ is the proton Coulomb potential. $H_0$ represents the unperturbed Hamiltonian and $H'$ is the perturbation term. We will disregard the electron spin in our treatment. 

\begin{figure}
\centering
\includegraphics[height=6cm]{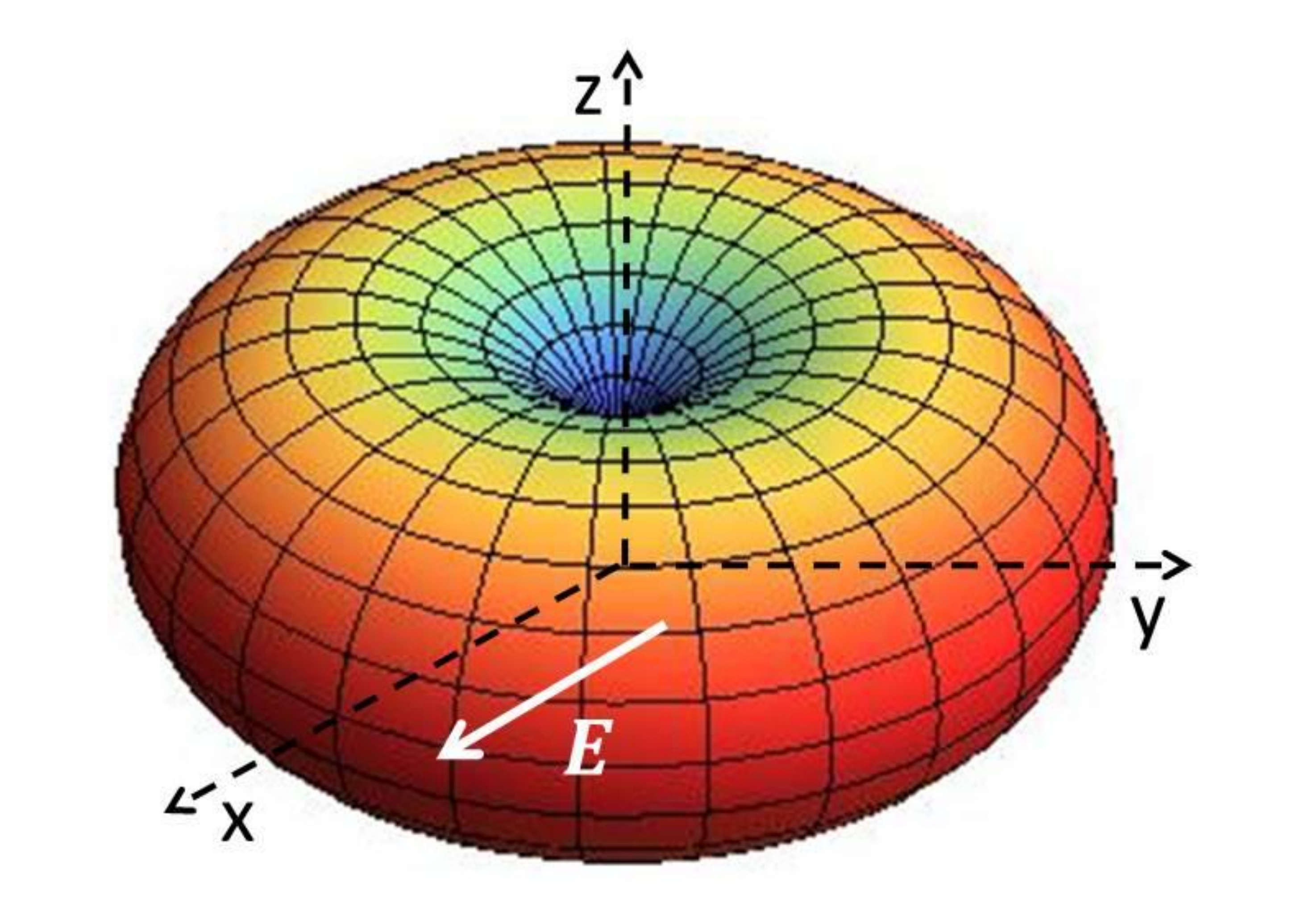}
\caption{The electronic cloud of a hydrogen atom initially in the unperturbed state $ \vert \psi^{(0)}_{2,1,1} \rangle$ has hidden momentum in the presence of a perturbation electric field $\mb{E}=E \mb{\hat{x}}$. The plot represents the surface with constant $|\psi|^2$, where $\psi$ is the wave function for this unperturbed state.}
\label{fig2}
\end{figure}

Consider the system initially in the unperturbed state $|\psi^{(0)}_{n,l,m}\rangle$, where $n$, $l$, and $m$ represent the principal quantum number, the orbital angular momentum quantum number, and the $z$ component of the orbital angular momentum in units of $\hbar$, respectively. For example, in the Fig. \ref{fig2} we have the plot of the surface with constant $|\psi|^2$, where $\psi$ is the wave function for the state $ \vert \psi^{(0)}_{2,1,1} \rangle$. Since this state has magnetic moment in the $-\mb{\hat{z}}$ direction, we expect that the atom should carry hidden momentum in the $-\mb{\hat{y}}$ direction when there is an external electric field in the $\mb{\hat{x}}$ direction filling all space, in analogy to the classical case. When this external electric field is present, the system state at time $t=0$ can be written as 
\begin{equation}
\label{state vector correction}
\vert {\psi_{n,l,m}} \rangle \approx |\psi^{(0)}_{n,l,m}\rangle+\sum \limits_{n'\neq n} \sum \limits_{l'} \sum \limits_{m'} C_{n',l',m'} \vert \psi_{n',l',m'}^{(0)} \rangle, 
\end{equation}
where $\vert \psi_{n',l',m'}^{(0)} \rangle$ are the eigenstates of the unperturbed Hamiltonian and $C_{n',l',m'}$ are small coefficients determined by perturbation theory \cite{mcintyrebook,bransdenbook,griffithsquantumbook}. In the Appendix we perform the explicit calculations for the state $\vert {\psi_{2,1,1}}\rangle$ using degenerate perturbation theory, justifying the form of Eq. (\ref{state vector correction}) at a particular instant of time, chosen to be $t=0$.  In the following, we calculate the hidden momentum for different states of the form of Eq. (\ref{state vector correction}) using two distinct methods. In the first one, we compute the expectation value of the relativistic momentum operator in the rest frame of the electronic cloud center-of-mass. In the second method, we use the quantum analogue of Eq. (\ref{hidden momentum babson}) transforming $\mb{J}$ and $\Phi$ into operators. By showing that the two methods provide close non-zero results for the hidden momentum of different quantum states, we prove its existence in this quantum system. We attribute the small differences between the values obtained by the two methods to the inaccuracy of the perturbation theory method \cite{griffithsquantumbook}. We consider the time $t=0$ in all our calculations.

The state (\ref{state vector correction}) is not an eigenstate of the Hamiltonian (\ref{total hamiltonian}), so the expectation values of many operators change in time. For instance, in the Appendix we show that for the state $\vert {\psi_{2,1,1}}\rangle$ the electronic cloud center-of-mass oscillates around the nucleus. The system hidden momentum is equal to the relativistic linear momentum when the classical linear momentum is null. For this reason, it is much easier to perform the calculations in the rest frame of the electronic cloud center-of-mass $\mathcal{S'}$.  But the nonrelativistic momentum operator $\mb{p}$ is defined in the proton rest frame $\mathcal{S}$, where the electronic cloud center-of-mass has a (non-relativistic) velocity in the $\mb{\hat{y}}$ direction given by $\left\langle \psi_{n,l,m}| \mb{p}|\psi_{n,l,m} \right\rangle/m_e\equiv v_c\mb{\hat{y}}$. The $x$ and $z$ components of this velocity were zero for all the quantum states we tested. Transforming the corresponding relativistic operator $\mb{P}=\gamma \mb{p}$ (with $\gamma=[1-{p}^{2}/(m_{e}^{2}c^{2})]^{-1/2}$) from reference $\mathcal{S}$ to $\mathcal{S'}$, the $y$ component is $P_{y}'  \approx \gamma p_{y} -  \gamma m_{e} v_{c}$ for $v_c\ll c$. Thus, considering only the terms up to second order in ${p}$ in the series expansion of $\gamma$, we obtain the $y$ component of the hidden momentum 
\begin{equation}
 P_{hid}'^{(1)} \equiv \frac{1}{2 m_{e}^{2}c^{2}} \left\langle \psi_{n,l,m} \vert \left( p_{y}-m_{e} v_{c} \right) {p}^{2}  \vert \psi_{n,l,m} \right\rangle,
\label{quantum hidden momentum 1}
\end{equation}
where  the superscript ``(1)" was used to label the hidden momentum calculated from the expected value of the relativistic linear momentum operator. The $x$ and $z$ components of the hidden momentum computed with this method were zero for all the quantum states we tested. In the following we use the superscript ``(2)" to label the hidden momentum calculated by the second method.

In our second method for computing the hidden momentum, we use the quantum analogue of Eq. (\ref{hidden momentum babson}). The current density $\mb{J}$ can be written as $\mb{J}=\rho\mb{v}=\rho\mb{p}/M$, where $\rho$ is a charge density, $\mb{v}$ is the velocity of this portion of charge, $\mb{p}$ its non-relativistic momentum and $M$ its mass. The charge density of the electronic cloud, on the other hand, can be associated to $-e|\psi_{n,l,m}(\mb{r})|^2$, where $\psi_{n,l,m}(\mb{r}) \equiv \langle \mb{r} \vert \psi_{n,l,m} \rangle$ is the electron wavefunction. So the quantum analogue of Eq.  (\ref{hidden momentum babson}) for the $y$ component of the hidden momentum is an expectation value of the form
\begin{equation}
P_{hid}'^{(2)} \equiv \frac{e}{2 m_{e} c^{2}} \int{ \psi_{n,l,m}^*(\mb{r'})  \left[ \Phi' p'_{y} + p'_{y} \Phi' \right]   \psi_{n,l,m}(\mb{r'}) d^3r'}. 
\label{quantum hidden momentum 2} 
\end{equation}
Again, the $x$ and $z$ components of the hidden momentum computed with this method were zero for all the quantum states we tested. The above parameters and operators are in the reference frame $\mathcal{S}'$, corresponding to the atomic cloud center-of-mass rest frame. Since the velocity $v_c\mb{\hat{y}}$ of this frame in relation to the reference frame $\mathcal{S}$ that correspond to the proton rest frame is non-relativistic, we can consider the transformations $\mb{r}'\approx\mb{r}$, $d^3r'\approx d^3r$, $\Phi' \approx  \Phi ={e}/({4 \pi \epsilon_{0} r})-E x$,  $p'_{y} \approx p_{y}-m_{e} v_{c}=-i \hbar {\partial}/{\partial y}-m_{e} v_{c}$. 

In order to provide a physical interpretation of Eq. (\ref{quantum hidden momentum 2}), it is convenient to rewrite it as a sum of two distinct terms
\begin{equation}
P_{hid}'^{(2)} = P_{hid}'^{(2a)}+P_{hid}'^{(2b)},
\label{quantum hidden momentum 2a+b} 
\end{equation}
with
\begin{equation}
P_{hid}'^{(2a)} \equiv - \frac{e E}{m_{e}c^{2}} \left\langle \psi_{n,l,m} \left| x p_{y}' \right| \psi_{n,l,m} \right\rangle\approx -\frac{m\mu_BE}{c^2},
\label{quantum hidden momentum 2a} 
\end{equation}
and
\begin{equation}
P_{hid}'^{(2b)} \equiv  \frac{ e}{8 \pi \epsilon_{0} c^{2}}  \left\langle \psi_{n,l,m} \left|  \left( \frac{1}{r} p_{y}' + p_{y}'\frac{1}{r}\right)  \right| \psi_{n,l,m} \right\rangle,
\label{quantum hidden momentum 2b} 
\end{equation}
where $\mu_{B} \equiv e \hbar/2m_{e}$ is the Bohr magneton. Considering terms up to the first order in $E$, $P_{hid}'^{(2a)}$ from Eq. (\ref{quantum hidden momentum 2a}) is exact and we can associate it to ${\boldsymbol{\mu}\times \mb{E}}/{c^{2}}$, since the magnetic moment of the unperturbed state is $\boldsymbol{\mu}=-m\mu_B\mb{\hat{z}}$. $P_{hid}'^{(2b)}$ is related to fact that the perturbed hydrogen atom has nonzero electric and magnetic dipole moments, which causes a second contribution to the system hidden momentum. 

Both methods for computing the system hidden momentum should be valid, so we expect to have $P_{hid}'^{(1)}=P_{hid}'^{(2)}$. According to Eqs. (\ref{quantum hidden momentum 1}), (\ref{quantum hidden momentum 2a+b}), (\ref{quantum hidden momentum 2a}), and (\ref{quantum hidden momentum 2b}) this equality can be written as 
\begin{equation}
 \frac{ P_{hid}'^{(1)} - P_{hid}'^{(2b)} }{\mu_{B}E/c^{2}} = -m.
 \label{graf}
\end{equation}
In order to check this claim, we computed numerically the values of $( P_{hid}'^{(1)} - P_{hid}'^{(2b)})c^{2}/(\mu_{B}E)$ for various quantum states. When computing the perturbed states, we considered their decomposition on the unperturbed states with $n'$ up to 20 in Eq. (\ref{state vector correction}). In the Fig. 3 we plot the values for this quantity for different quantum states with the quantum number $m$ varying from -5 to 5. The blue line is the prediction from Eq. (\ref{graf}). It can be seen that there is a good agreement between the hidden momentum values obtained with the two methods. We attribute the small differences to the inaccuracy of the perturbation theory method in determining the perturbed wavefunctions \cite{griffithsquantumbook}.
 
\begin{figure}
\centering
\includegraphics[height=6cm]{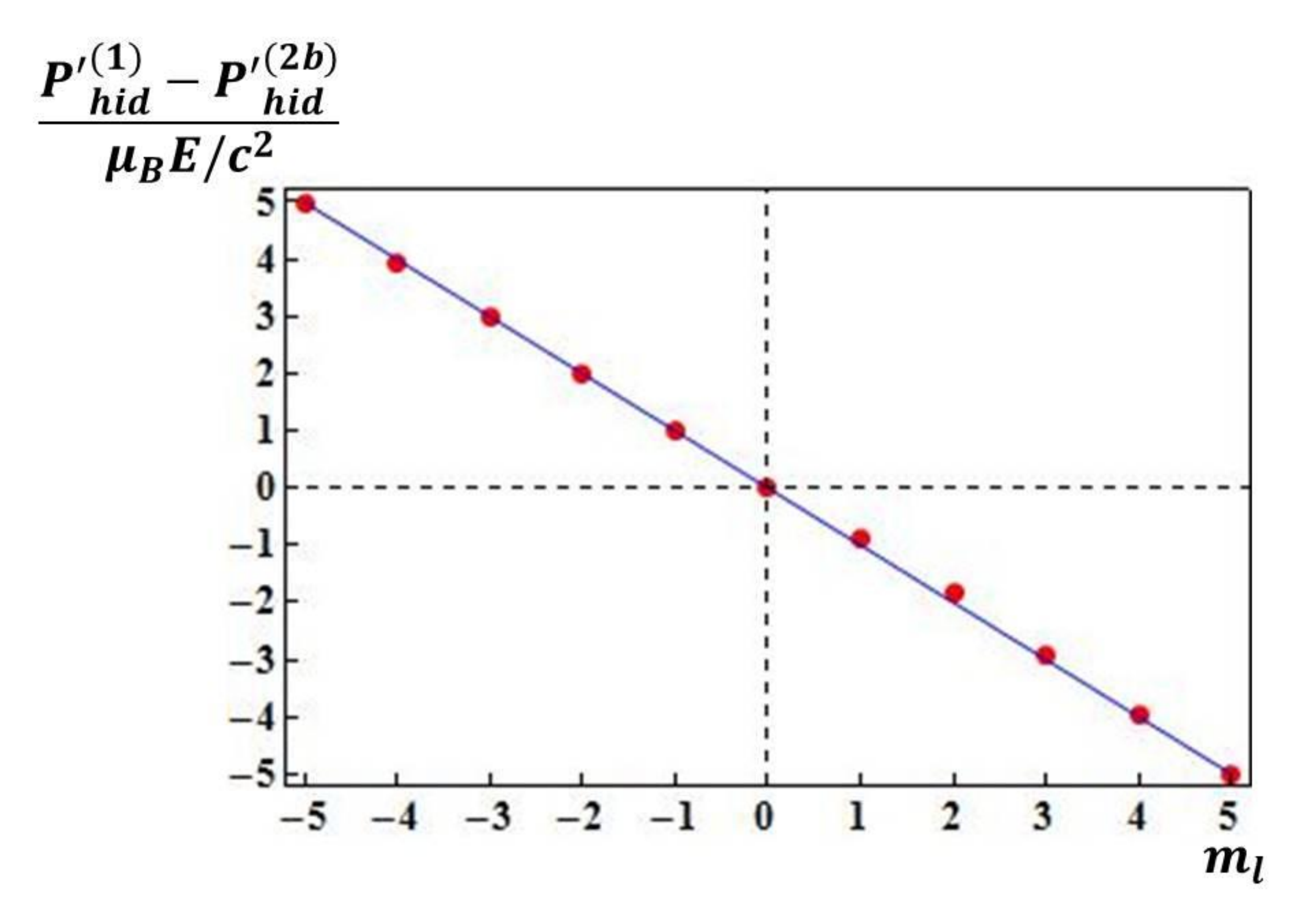}
\caption{The blue straight line is $y=-m_{l}$ and the red dots represent the values of ${\left( P_{hid}'^{(1)} - P_{hid}'^{(2b)} \right)}c^{2}/{\mu_{B}E}$ for the following quantum states $\vert {\psi_{n,l,m}} \rangle$: $\left\lbrace n,l,m\right\rbrace =$ $\left\lbrace 13,12,-5 \right\rbrace ,$ $\left\lbrace 11,7,-4 \right\rbrace,$ $\left\lbrace 6,5,-3 \right\rbrace,$  $\left\lbrace 12,8,-2 \right\rbrace,$ $\left\lbrace 3,1,-1 \right\rbrace,$ $\left\lbrace 1,0,0 \right\rbrace,$ $\left\lbrace 2,1,1 \right\rbrace,$ $\left\lbrace 9,2,2 \right\rbrace,$ $\left\lbrace 8,4,3 \right\rbrace,$ $\left\lbrace 5,4,4 \right\rbrace,$ $\left\lbrace 7,6,5 \right\rbrace$.}
\label{graph1}
\end{figure}

Now we consider the hydrogen atom in the state $\vert \psi_{3,1,-1} \rangle$ in the presence of an external electric field given by $\mb{E} =E \left(\hat{\mb{x}} \cos{\theta} + \hat{\mb{z}} \sin{\theta}  \right) $.  In this situation we have $P_{hid}'^{(2a)}={\mu_BE}/{c^2}\cos(\theta)$ up to the first order in $E$, which again can be associated to ${\boldsymbol{\mu}\times \mb{E}}/{c^{2}}$. The $x$ and $z$ components of the hidden momentum are zero. In the Fig. 4 the red dots represent the numerical values obtained for ${( P_{hid}'^{(1)} - P_{hid}'^{(2b)})}c^{2}/({\mu_{B}E})$ for different angles $\theta$ and the blue curve corresponds the expected expression $\cos(\theta)$. Again, there is a good agreement.

\begin{figure}
\centering
\includegraphics[height=6cm]{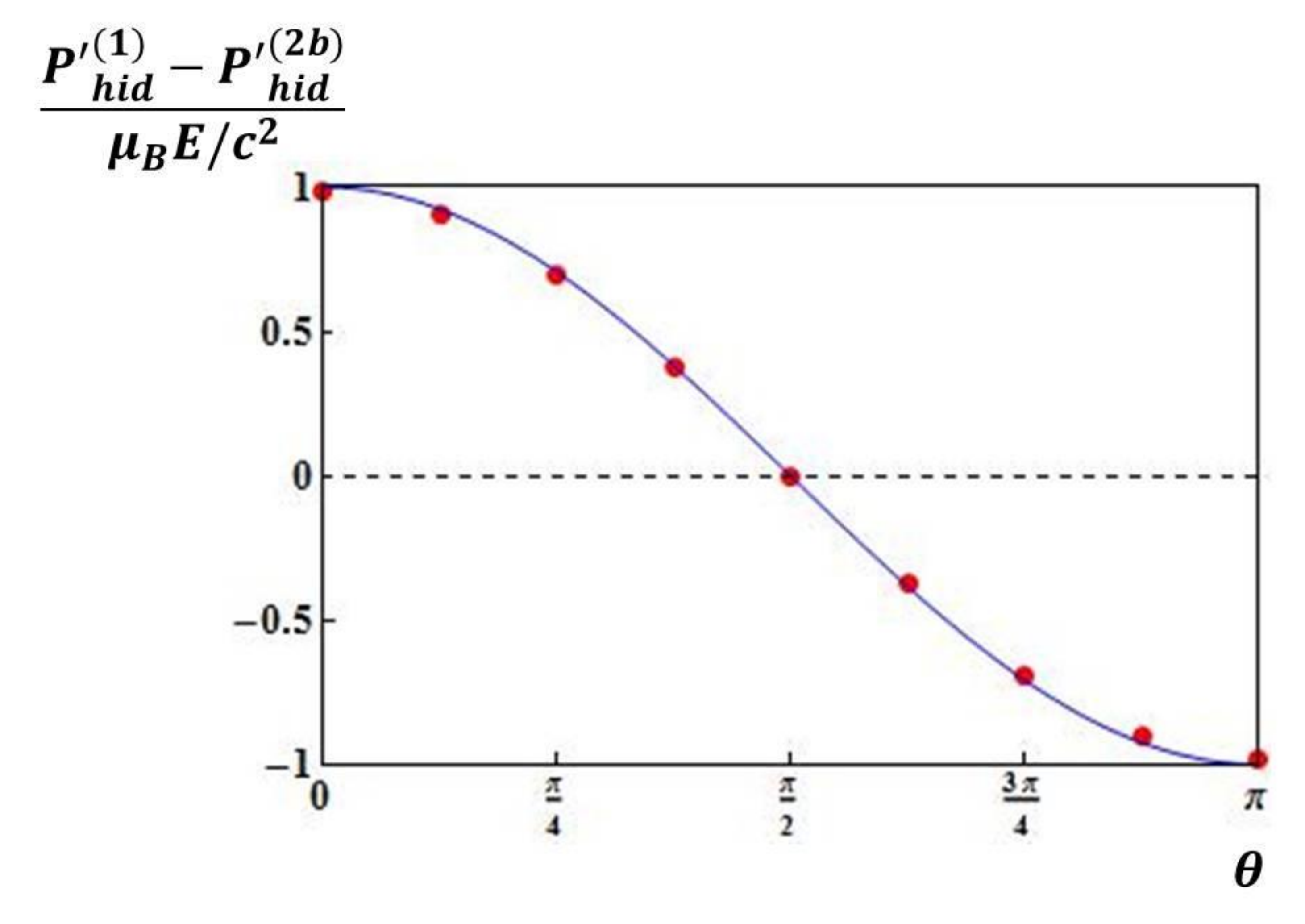}
\caption{The blue line is $y=\cos{\theta} $ and the red dots represent the values of ${\left( P_{hid}'^{(1)} - P_{hid}'^{(2b)} \right)}c^{2}/{\mu_{B}E}$ for the quantum state $\vert \psi_{3,1,-1} \rangle$ perturbed by an electric field $\mb{E} =E \left(\hat{\mb{x}} \cos{\theta} + \hat{\mb{z}} \sin{\theta}  \right) $ for different values of $\theta$.}
\label{graph2}
\end{figure}

The results expressed in the graphs of Figs. 3 and 4 attest to the existence of hidden momentum in the system, as required to guarantee the validity of the Lorentz force law in the quantum limit. This means that the equations for the force and torque on a quantum magnetic dipole with dipole moment $\boldsymbol{\mu}$ due to the orbital angular momentum of quantum particles has the same form as the force and torque on a classical magnetic dipole with the same magnetic dipole moment $\boldsymbol{\mu}$.

It is important to make it clear that so far we have taken into account only the electron orbital angular momentum, disregarding the electron spin, to compute the atom magnetic moment and the hidden momentum in the system. But does an electron with magnetic moment resulting from its spin in the presence of an applied electric field have hidden momentum? We cannot answer this question using the method we have used so far, since there is no physical model for the origin of spin in terms of the movement of constituent parts of the electron. Even if we treat the problem using quantum field theory the question cannot be solved, since the electron is a fundamental particle with no constituent parts, so we cannot deduce if it has or if it has not hidden momentum in the presence of an external field. We then have two options: we can postulate that there is hidden momentum for spin or we can postulate that there is no hidden momentum for spin. The problem on postulating the non-existence of hidden momentum for spin is that, in order to avoid paradoxes, we would need two force equations in this case: one to the part of the magnetic moment that is due to spin and another (the Lorentz force) to the part of the magnetic moment that is due to the orbital angular momentum of the electron. This would be a weird choice, but the present work cannot deny this possibility. A much simpler alternative is to postulate the existence of hidden momentum for spin in the presence of an external electric field. In this way the Lorentz force law should be valid in generic quantum systems.

\section{Abraham-Minkowski debate}

The hidden momentum concept is related to a long-standing discussion, known as the Abraham-Minkowski debate, about the expression for the momentum density of electromagnetic waves in material media \cite{pfeifer1}. In the Abraham (Minkowski) formulation, this momentum density assumes the form $\mb{E} \times \mb{H}/c^{2}$ ($\mb{D} \times \mb{B}$), where $\mb{E}$ is the electric field, $\mb{B}$ is the magnetic field, $\mb{D}\equiv \varepsilon_0 \mb{E}+\mb{P}$ is the electric displacement, and $ \mb{H}\equiv \mb{B}/\mu_o-\mb{M}$, with $\mb{P}$ and $\mb{M}$ being the polarization and magnetization of the medium. The eventual conclusion of this debate is that there are several ways for dividing the total energy-momentum tensor of the system in two parts, the electromagnetic and the material ones, which lead to the same experimental predictions for all measurable physical quantities \cite{penfieldbook,pfeifer1}. In this sense both Abraham and Minkowski formulations are incomplete, since they consider only the electromagnetic energy-momentum tensor, disregarding the material energy-momentum tensor. When the appropriate material tensors are taken into account for each formulation, the experimental predictions are the same \cite{penfieldbook,pfeifer1}. In a recent work, Barnett has shown that we can associate the Abraham and Minkowski momenta with the kinetic and canonical momenta of light, respectively \cite{barnett2} (see also \cite{barnett3}). It was also recently shown by one of us that when the Lorentz force is used to compute the momentum transfer from the electromagnetic wave to the material medium, the electromagnetic part of the momentum density must be $\epsilon_{0} \mb{E} \times \mb{B}$ in order to obtain momentum conservation in several situations \cite{saldanha2,saldanha3}. In nonmagnetic media, $\epsilon_{0} \mb{E} \times \mb{B}$ is equal to the Abraham momentum. But in magnetized media there is a hidden momentum density $\mb{M} \times \mb{E}/c^{2}$, compatible with the use of the Lorentz force law, which is not associated to the motion of the material constituents, thus not being a kinetic momentum. When we consider only the material kinetic momentum, $\mb{M} \times \mb{E}/c^{2}$ must be considered as part of the electromagnetic momentum density, that becomes $\epsilon_{0} \mb{E} \times \mb{B}+\mb{M} \times \mb{E}/c^{2}=\mb{E} \times \mb{H}/c^{2}$, which is the Abraham momentum. So the existence of hidden momentum in the medium and the use of the Lorentz force law are consistent with the association of the Abraham momentum to the kinetic momentum of light, and consequently also to the association of the Minkowski momentum to the canonical momentum of light. 

\section{Conclusion}

In this work we used perturbation theory to compute the hidden momentum of a hydrogen atom in the presence of an external electric field when the magnetic dipole moment is due to the orbital angular momentum of the electron. We used two different methods for computing this quantity and obtained the same results, evidencing the existence of hidden momentum in the system and the consequent validity of the Lorentz force law in quantum systems. This fact makes $\epsilon_{0} \mb{E} \times \mb{B}$ a valuable expression for the electromagnetic momentum of light in a medium, since it is compatible with the use of the Lorentz force law in computing the momentum transfer from light to the material medium. 

\section*{ Acknowledgements}
We acknowledge Mario Mazzoni and David J. Griffiths for useful discussions. This work was supported by the Brazilian agencies CNPq and CAPES.

\appendix*

\section{ }

In this appendix we use first order degenerate perturbation theory to compute the perturbed state $\vert \psi_{2,1,1} (t) \rangle$ with the Hamiltonian (\ref{total hamiltonian}). Since we are not considering the fine and hyperfine structure of hydrogen, the unperturbed states $\vert \psi_{2,0,0}^{(0)} \rangle$, $\vert \psi_{2,1,-1}^{(0)} \rangle$, $\vert \psi_{2,1,0}^{(0)} \rangle$ and $\vert \psi_{2,1,1}^{(0)} \rangle$ have the same energy $E_2^{(0)}$, such that degenerate perturbation theory must be used \cite{mcintyrebook,bransdenbook,griffithsquantumbook}. The normalized eigenvectors obtained by diagonalizing the operator $H'$ in the subspace spanned by these unperturbed states are
\begin{equation}
\vert \eta_{1}^{(0)} \rangle = \frac{1}{2} \left[ -\sqrt{2} \vert \psi_{2,0,0}^{(0)} \rangle - \vert \psi_{2,1,-1}^{(0)} \rangle + \vert \psi_{2,1,1}^{(0)} \rangle \right],  
\label{a11}
\end{equation}
\begin{equation}
\vert \eta_{2}^{(0)} \rangle = \frac{1}{2} \left[ \sqrt{2} \vert \psi_{2,0,0}^{(0)} \rangle - \vert \psi_{2,1,-1}^{(0)} \rangle + \vert \psi_{2,1,1}^{(0)} \rangle \right], 
\label{a12}
\end{equation}
\begin{equation}
\vert \eta_{3}^{(0)} \rangle = \frac{1}{\sqrt{2}} \left[ \vert \psi_{2,1,-1}^{(0)} \rangle + \vert \psi_{2,1,1}^{(0)} \rangle \right], 
\label{a13}
\end{equation}
\begin{equation}
\vert \eta_{4}^{(0)} \rangle = \vert \psi_{2,1,0}^{(0)} \rangle, 
\label{a14}
\end{equation}
and their respective eigenvalues are $E_{\eta_{1}}^{(1)}=-3 a_{0} e E$, $E_{\eta_{2}}^{(1)}=3 a_{0} e E$, $E_{\eta_{3}}^{(1)}=0$ and $E_{\eta_{4}}^{(1)}=0$, where $a_0$ is the Bohr radius. Since we have $\vert \psi_{2,1,1}^{(0)} \rangle=\frac{1}{2}\vert\eta_{1}^{(0)}\rangle+\frac{1}{2}\vert\eta_{2}^{(0)}\rangle+\frac{1}{\sqrt{2}}\vert\eta_{3}^{(0)}\rangle$, the perturbed state $\vert \psi_{2,1,1}(t) \rangle$ at time $t$ is
\begin{eqnarray}\label{a17}
&&\vert \psi_{2,1,1}(t) \rangle \approx \frac{1}{2} \left[ \vert \eta_{1}^{(0)} \rangle + \vert \eta_{1}^{(1)} \rangle \right] \mathrm{e}^{{3 i a_{0} e E t}/{\hbar}} + \\ \nonumber
&&+\frac{1}{2} \left[\vert \eta_{2}^{(0)} \rangle + \vert \eta_{2}^{(1)} \rangle \right] \mathrm{e}^{{-3 i a_{0} e E t}/{\hbar}} + \frac{1}{\sqrt{2}} \left[ \vert \eta_{3}^{(0)} \rangle + \vert \eta_{3}^{(1)} \rangle \right],
\end{eqnarray}
where the global phase $\mathrm{e}^{-i E_{2}^{(0)} t/\hbar}$ was omitted and the first order correction terms are
\begin{equation}
\vert \eta_{\alpha}^{(1)} \rangle = \sum \limits_{n' \neq 2} \sum \limits_{l'} \sum \limits_{m'} \frac{\langle \psi_{n',l',m'}^{(0)} \vert H' \vert \eta_{\alpha}^{(0)} \rangle}{E_{2}^{(0)}-E_{n'}^{(0)}} \vert \psi_{n',l',m'}^{(0)} \rangle.
\label{a16}
\end{equation}

With the state $\vert \psi_{2,1,1}(t) \rangle$ from Eq. (\ref{a17}), we compute the expected value of the $y$ component of the electron position (without the first order correction terms from Eq. (\ref{a16})): 
\begin{equation}
\langle y \rangle (t) \approx - 3 a_{0} \sin \left[  \frac{3 a_{0} e E t}{\hbar} \right].
\label{a19}
\end{equation}
We see that the electronic cloud center-of-mass oscillates around the nucleus. Performing a numerical calculation with the perturbed state considering correction terms up to $n'=20$ in Eq. (\ref{a16}), we obtain the expectation value of the $y$ component of the electron momentum:
\begin{equation}
\langle p_{y} \rangle (t) \approx  - 8.25\frac{a_{0}^2eEm_e}{\hbar}   \cos  \left[  \frac{3 a_{0} e E t}{\hbar} \right].
\label{a22}
\end{equation}
We can see that $m_{e} \frac{d \langle y \rangle (t)  }{dt} \approx 1.09 \langle p_{y} \rangle (t) $. This difference between the values can be attributed to the inaccuracy of the perturbation theory method, which is not very precise for determining the perturbed states \cite{griffithsquantumbook}.

At time $t=0$, the perturbed state of Eq. (\ref{a17}) can be written as
\begin{equation}
\vert {\psi_{2,1,1}} \rangle \approx |\psi^{(0)}_{2,1,1}\rangle+\sum \limits_{n'\neq 2} \sum \limits_{l'} \sum \limits_{m'} C_{n',l',m'} \vert \psi_{n',l',m'}^{(0)} \rangle, 
\end{equation}
with
\begin{equation}
 C_{n',l',m'}=\frac{\langle \psi_{n',l',m'}^{(0)} \vert H' \vert \psi^{(0)}_{2,1,1}\rangle}{E_{2}^{(0)}-E_{n'}^{(0)}} \vert \psi_{n',l',m'}^{(0)} \rangle.
\end{equation}
Note that if we are interested on the perturbed state at time $t=0$, a diagonalization of the perturbation term $H'$ is not necessary. This is also true for the other perturbed states that we consider. For this reason in our work we use the expression (\ref{state vector correction}) for the perturbed states to compute the system hidden momentum at time $t=0$.

\end{document}